\documentclass[12pt,preprint]{aastex} 
\usepackage{psfig,lscape}
\def\lap{\lower.5ex\hbox{$\; \buildrel < \over \sim \;$}}
\def\gap{\lower.5ex\hbox{$\; \buildrel > \over \sim \;$}}

\def\ergcm2s{${\rm erg\ cm^{-2}\ s^{-1}}$}

\def\ergscm2s{${\rm erg\ cm^{-2}\  s^{-1}}$}

\def\cm-2{${\rm cm^{-2}}$}

\def \eg{{e.g.,}}

\def \ie{{i.e.,}}

\begin{document}

\title{The Metallicity Distribution of Intracluster Stars in Virgo}

\author{Benjamin F. Williams, Robin Ciardullo\altaffilmark{1}, Patrick
R. Durrell\altaffilmark{2}, Matt Vinciguerra\altaffilmark{1}, John
J. Feldmeier\altaffilmark{2}, George H. Jacoby\altaffilmark{3}, Steinn
Sigurdsson \altaffilmark{1}, Ted von Hippel\altaffilmark{4}, Henry
C. Ferguson\altaffilmark{5}, Nial R. Tanvir\altaffilmark{6}, Magda
Arnaboldi\altaffilmark{7}, Ortwin Gerhard\altaffilmark{8}, J. Alfonso
L. Aguerri\altaffilmark{9}, and Ken Freeman \altaffilmark{10}}

\altaffiltext{1}{Department of Astronomy \& Astrophysics, 
Pennsylvania State University, University Park, PA 16802; 
bwilliams@astro.psu.edu; rbc@astro.psu.edu; steinn@astro.psu.edu}
\altaffiltext{2}{Department of Physics \& Astronomy, Youngstown State 
University, Youngstown, OH 44555; prdurrell@ysu.edu; jjfeldmeier@ysu.edu NSF Astronomy and Astrophysics Postdoctoral Fellow}
\altaffiltext{3}{WIYN Observatory, 950 North Cherry Avenue, P.O. Box 26732, 
Tucson, AZ 85726; jacoby@wiyn.org}
\altaffiltext{4}{The University of Texas, Department of Astronomy, 
1 University Station C1400, Austin, Texas 78712; ted@astro.as.utexas.edu}
\altaffiltext{5}{Space Telescope Science Institute, 3700 San Martin Drive, 
Baltimore, MD 21218; ferguson@stsci.edu}
\altaffiltext{6}{University of Hertfordshire, College Lane, Hatfield, 
Herts AL10 9AB, UK; nrt@star.herts.ac.uk}
\altaffiltext{7}{European Southern Observatory, Karl-Schwarzchild-Str.~2,
85748 Garching, Germany, marnabol@eso.org}
\altaffiltext{8}{Max-Planck-Institut fuer Extraterrestrische Physik,
P.O. Box 1312, D-85741 Garching, Germany, gerhard@exgal.mpe.mpg.de}
\altaffiltext{9}{Instituto de Astrofisica de Canarias, C/ V\'{i}a L\'{a}ctea, 
s/n, 38205, La Laguna, Tenerife, Spain; jalfonso@ll.iac.es}
\altaffiltext{10}{Mount Stromlo Observatory, Research School of Astronomy 
and Astrophysics, Mount Stromlo Observatory, The Australian National 
University, ACT 0200 Australia; kcf@mso.anu.edu.au}

\keywords{ galaxies: clusters: individual (Virgo) ---  galaxies: evolution}

\begin{abstract}
We have used the {\sl Hubble Space Telescope's Advanced Camera for
Surveys (ACS)\/} to detect and measure $\sim 5300$ stars in a single
intracluster field in the Virgo Cluster.  By performing F606W and
F814W photometry on these stars, we have determined their metallicity
distribution function, and constrained the types of stars present in
this portion of Virgo's intracluster space.  Based on the small number
of stars detected brighter than the red giant branch (RGB) tip, we
suggest that in this region, Virgo's intracluster stars are mostly old
($\gap$10~Gyr).  Through analysis of the RGB stars themselves, we
determine that the population contains the full range of metallicities
probed ($-2.3 \leq$~[M/H]~$\leq 0.0$).  We also present evidence that
the younger ($\leq$10~Gyr) component of the population is more
metal-rich, with [M/H]~$> -0.5$.  The spatial distribution of the most
metal-poor stars in the field shows significantly more structure than
that of the metal-rich stars, indicating that the intracluster
population is not well-mixed.  We discuss the implications these
observations have for the production of intracluster stars and the
dynamical evolution of the Virgo Cluster.

\end{abstract}

\section{Introduction}

Galaxy interactions and mergers are fundamental to the process of
galaxy evolution.  During these encounters, tidal forces often eject
stars from their parent galaxies into intergalactic space; these
orphaned stars then provide a fossil record of the interaction.  By
studying the chemistry and age distribution of these castaway objects,
we can constrain models for the galaxies' formation and evolution
\citep[\eg][]{napolitano2003, murante2004, willman2004, sommer2005}.
For example, the metallicities of the ejected stars give us clues
about the kind of galaxies from which they originate.  Likewise, the
number and spatial distribution of the stars lend insight about the
types of interactions involved and their frequency, and the kinematics
and phase-space structure of the stars provide details about the
history of a group or cluster.  As a result, studies of orphaned red
giants, planetary nebulae, and globular clusters can directly
constrain the interaction history of a system of galaxies.

The intracluster medium is particularly suitable for the measurement
of intergalactic stars.  Galaxies in dense regions of space are known
to have redder colors and possess less gas than their field
counterparts \citep[\eg][]{dressler1980}.  Similarly, cluster galaxies
appear to have evolved more rapidly than field galaxies, as exhibited
by their observed color shift to the red over the past few billion
years \citep{butcher1978}.  There are many possible explanations for
these differences, including gravitational interactions and mergers
among cluster galaxies \citep[\eg][]{toomre1972, richstone1976,
moore1996, dubinski1998,rudick2006}, gravitational interactions with
the cluster potential \citep[\eg][]{byrd1990, moore1998, gnedin2003},
and the loss of gas via interactions with the intracluster medium
\citep[\eg][\ and many others]{gunn1972, larson1980, fujita1999,
bekki2002}.  Each of these processes can leave noticeable imprints on
the population of stars that reside between the galaxy cluster
members.

There have been several numerical simulations of the intracluster
light phenomenon.  Models by \citet{murante2004}, focusing on
117 clusters with masses ranging from $10^{14} M_{\odot}$ to
$10^{15} M_{\odot}$, suggest that between 10\% and 50\% of a
system's stars may be unattached to any galaxy, with the more massive
clusters having the larger fraction of intracluster stars.
Their findings also predict that the intracluster population should be
older and more centrally condensed than the galactic stellar
population.  Alternatively, the simulations of \citet{sommer2005} of 
one Virgo-like cluster and one Coma-like cluster produce a range for 
the percentage of intracluster stars (20\% to 40\%) and yield higher 
mean stellar metallicities, which decrease from near solar at the 
cluster core, to half-solar at the virial radius.  Still another set of
higher-resolution models by \citet{willman2004} predicts an
intracluster star fraction of between 10\% and 22\% in a rich galaxy
cluster, with most of the objects coming from the outer, metal-poor
regions of galaxies.  The \citet{willman2004} models also suggest that
massive galaxies are an important contributor to the intracluster
stellar population, and that the mean metallicity of these orphaned
stars is [M/H] $\gtrsim -1$.

Despite these theoretical advances, direct measurements of the
intracluster population have been difficult.  Although the existence
of intracluster light has been known for some time \citep[see][\ for
early studies]{zwicky1951, oemler1973, mattila1977}, only recently has
the importance of this component been quantified across different
cluster environments \citep[][and references therein]{ciardullo2004,
zibetti2005, feldmeier2004b, aguerri2005, gonzalez2005, krick2006}.
Moreover, little is known about the stellar populations that make up
the intracluster light.  Most analyses assume that the component is
old, but the lone observational constraint on the population's
metallicity comes from {\sl HST\/} $I$-band photometry of stars at the
tip of the red giant branch.  By comparing the brightnesses of these
stars to that of RGB tip stars in a metal-poor dwarf galaxy,
\citet{durrell2002} estimated the mean metallicity of the component to
be $-0.8 \lesssim {\rm [M/H]} \lesssim -0.2$.  Unfortunately, since
this estimate assumes that the dwarf galaxy and the intracluster stars
are at exactly the same distance, it is subject to a large systematic
error.

The best place to further our knowledge of intracluster light is in
the Virgo cluster.  At a distance of $\sim 15$~Mpc
\citep{freedman2001}, Virgo is the nearest system to have a
significant intracluster population \citep{ciardullo2004}, and it is
thus the best studied system to date.  There have been numerous
surveys of the cluster's intracluster planetary nebulae
\citep{arnaboldi1996, arnaboldi2002, feldmeier2004} which have helped
trace the distribution of the intracluster stars \citep{arnaboldi2003,
arnaboldi2004, aguerri2005}, and two small intracluster fields have
been imaged with the {\sl WFPC2\/} instrument of the {\sl Hubble Space
Telescope\/} \citep{ferguson1998, durrell2002}.  In addition, the
large-scale structure of Virgo's intracluster light has been mapped to
a very low level via the ultra-deep surface photometry of
\citet{mihos2005}.

Here, we present the results of a deep {\sl HST\/} photometric survey
of an intracluster field of the Virgo Cluster.  We use the {\sl
Advanced Camera for Surveys (ACS)\/} to create an F606W-F814W
color-magnitude diagram (CMD) of the intracluster stars and use the
data to constrain the age and metallicity of the stellar population.
In Section~2, we describe our survey and the analysis techniques
required to measure the system's metallicity distribution function.
In Section~3, we discuss the results of our analysis, and their
interpretation.  We show that Virgo's intracluster stars are mostly
old ($>10$~Gyr), and metal-poor, with a metallicity distribution
function that contains the full range of metallicities probed by the
data.  However, we also present evidence for the existence of a
younger, more metal-rich component, which may represent the results of
on-going tidal processes.  Finally, in Section~4, we discuss how these
results can be used to test models of cluster formation and evolution.

\section{Data Acquisition and Reduction}

\begin{figure}
\centerline{\psfig{file=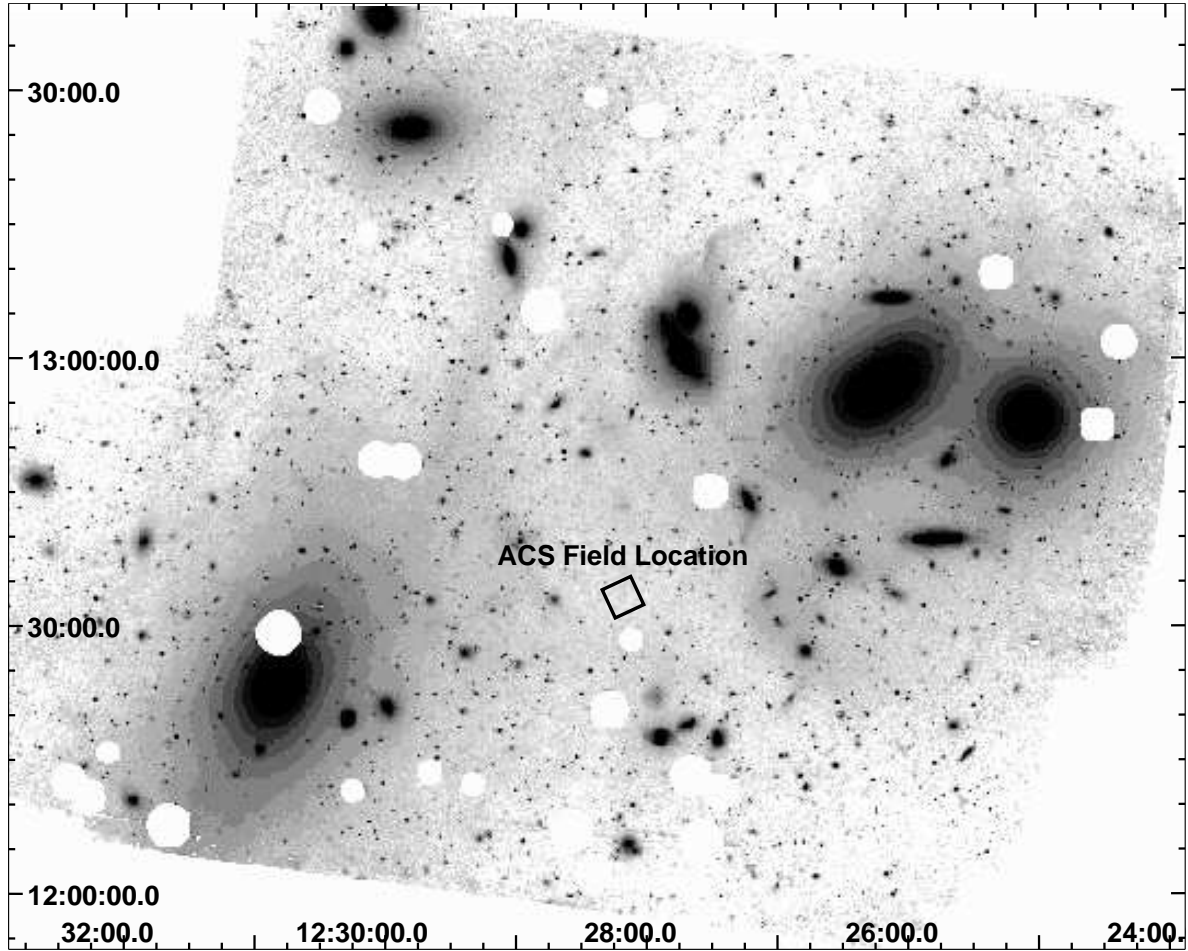,height=5.0in,angle=0}}
\caption{A deep ($\mu_V<28$ mag arcsec$^{-2}$) image of the Virgo Cluster
core \citep{mihos2005}, with the outline of our intracluster field 
superposed.  At the distance of Virgo, this field is over 
170~kpc from the nearest large galaxy.}
\label{mihos}
\end{figure}

From 30 May 2005 to 7 June 2005, we obtained deep {\sl HST ACS\/}
images of a Virgo Cluster intracluster field as part of project GO
10131.  Our single field, located at $\alpha(2000) = $ 12:28:10.80,
$\delta(2000) = $ 12:33:20.0 (with an orientation of 112.58 degrees),
was chosen to avoid any known galaxy in the Virgo Cluster Catalog,
filaments, arcs, or other low-surface brightness features of the
system.  Figure~\ref{mihos} shows the location of our field superposed
on the ultra-low surface brightness image of \citet{mihos2005}.  As
illustrated in the figure, our field is $40\arcmin$ (190~kpc) from
M87, $37\arcmin$ (170~kpc) from M86, and located on a patch of sky
which has a mean $V$-band surface brightness (including background
galaxies) of $\mu \sim 27.7$ \citep{mihos2005}.

\begin{figure}
\centerline{\psfig{file=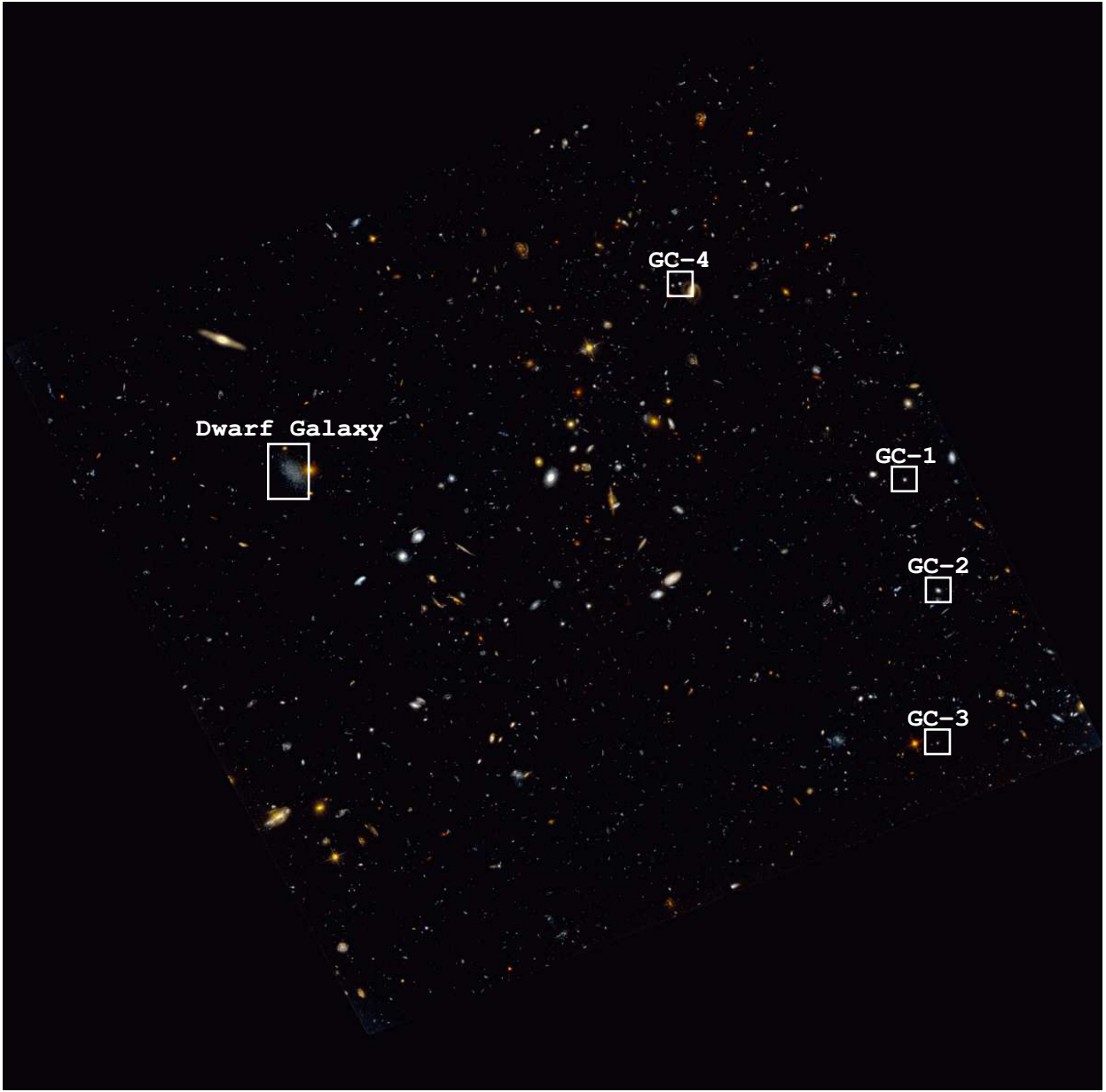,width=6.5in,angle=0}}
\caption{An image of our {\sl ACS\/} survey field.  In the image, blue
represents $2\times$ F606W$-$F814W, green represents F606W, and red
represents F814W.  The image is oriented North-up, East-left, and the
field of view is 202~arcsec on a side.  Contained in the image are
four candidate intracluster globular clusters, one previously
undiscovered Virgo Cluster dwarf spheroidal galaxy, numerous
background galaxies, and $\sim 5300$ intracluster stars.}
\label{fullimage}
\end{figure}

\begin{figure}
\centerline{\psfig{file=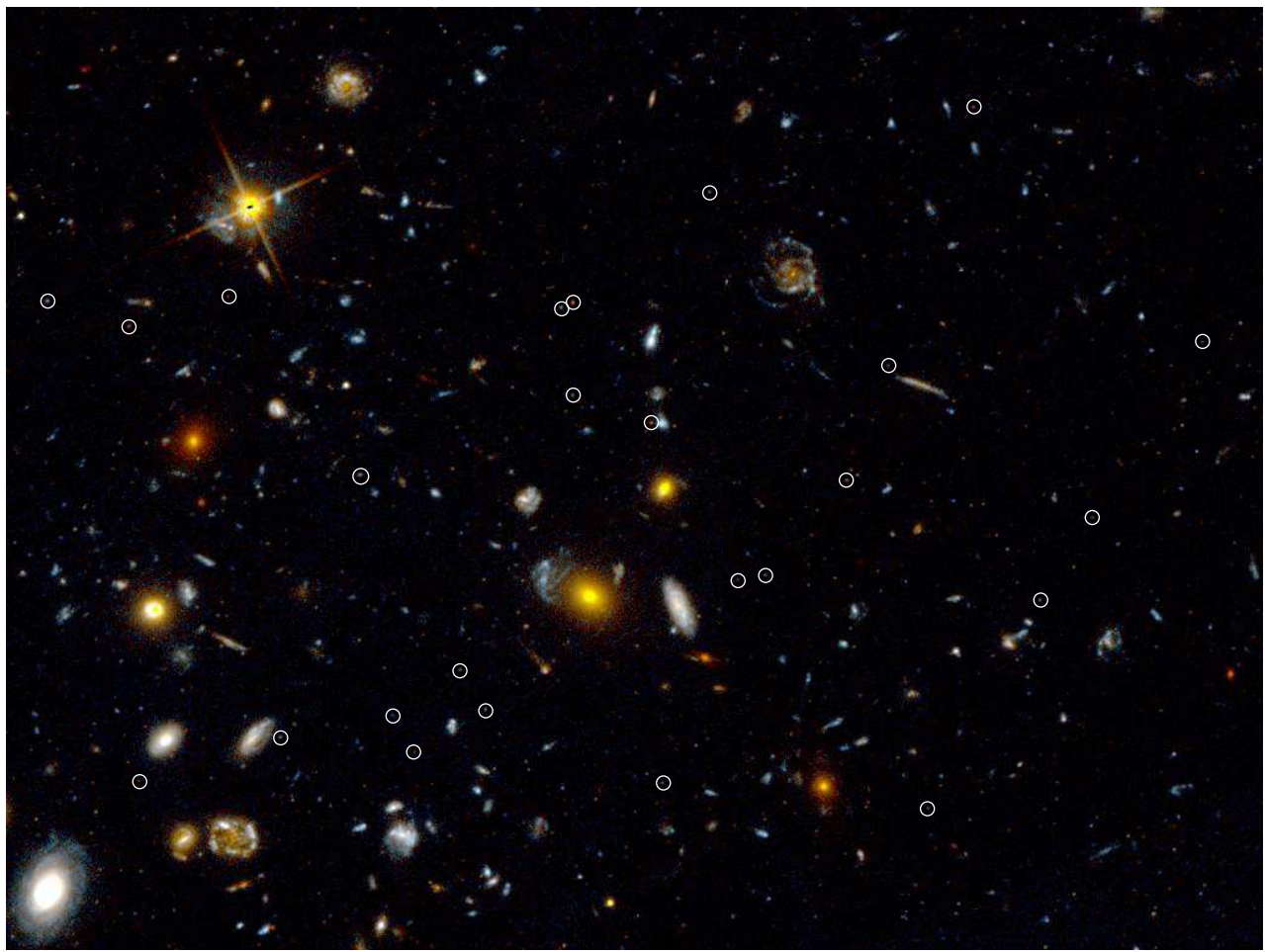,height=5.0in,angle=0}}
\caption{A $1\farcm 0 \times 0\farcm 8$ portion of our {\sl ACS\/}
field.  In the image, blue represents $2\times$F606W$-$F814W, green
represents F606W, and red represents F814W.  White circles indicate
the positions of some objects that match the image PSF and are probable
intracluster stars.}
\label{field}
\end{figure}

Observations were taken through two filters, the wide $V$-band (F606W)
and the standard $I$ (F814W).  The F814W data were taken over a period
of 11 orbits and consisted of 22 exposures totaling 26880 s of
exposure time; the F606W frames were acquired over 26 orbits, and
included 52 exposures with 63440 s of integration.  These frames were
combined using the PyRAF\footnote{PyRAF is a product of the Space
Telescope Science Institute, which is operated by AURA for NASA.} task
{\tt multidrizzle},\footnote{multidrizzle is a product of the Space
Telescope Science Institute, which is operated by AURA for
NASA. http://stsdas.stsci.edu/pydrizzle/multidrizzle} which removed
cosmic ray events and geometric distortions from our series of
dithered images.  The result was two co-added photometric images (in
units of counts per second), covering an exposed area of 11.39
arcmin$^2$ with a sampling almost twice that of the physical {\sl
ACS\/} detector, or $0 \farcs 03$~pixel$^{-1}$.  Unfortunately, the
point spread functions (PSFs) of our co-added F606W and F814W images
were not identical.  Some of the F606W frames had image headers whose
coordinate systems were mis-aligned by $\sim$1 pixel.  We corrected
for these misalignments using the positions of point sources in the
field, but due to the small number of bright targets available for
centroiding, the accuracy of this procedure was limited to $\sim 0
\farcs 02$.  This caused a slight degradation in the F606W PSF
compared to that of the F814W image.  A color reproduction of our
final image is shown in Figure~\ref{fullimage}; a portion of this
image is shown in color in Figure~\ref{field}.  A handful of the
sources that fit the image PSF and have magnitudes consistent with
those of Virgo Cluster red giant stars are circled.

A cursory visual examination of our {\sl ACS\/} images reveals several
interesting objects.  In the northeast region of our field is a
previously undiscovered Virgo Cluster low-surface brightness dwarf
spheroidal galaxy \citep{VICS3}.  Also present in the frame are four
candidate intracluster globular clusters \citep{VICS2}.  Finally, our
field contains a large number of background galaxies, both resolved and
unresolved.  Though these galaxies may be the focus of a future study,
here they are only a contaminant.  Nevertheless, we need to understand
this component if we are to extract the maximum amount of information
from the intracluster stars.

\subsection{Point Spread Function Photometry}

We performed PSF photometry on all objects in our field using the {\tt
DAOPHOT II} and {\tt ALLSTAR} packages \citep{stetson1990}.  We first
used {\tt DAOFIND} to create a list of sources by searching the image
for peaks that were 3$\sigma$ above the mean in each band.  The
brightest of these sources were then examined by eye, in order to
identify a set of objects which could be used to defined the image
point spread function (49 such sources were found, allowing us to fit
for a varying PSF across the field of view).  Once this was done,
{\tt ALLSTAR} was used to fit this PSF to the remaining 3$\sigma$
peaks; this generated two lists of magnitudes and fitting statistics,
one for each filter.  These lists were then cross-correlated, to
produce a catalog of $\sim 11,000$ sources present on both frames.

We next attempted to remove background galaxies from our source catalog by 
eliminating those objects with DAOPHOT $\chi^2$ fits to the PSF larger than 
1.25, absolute roundness values greater than 1.0, or image centroids that 
differed by more than one pixel from one image to the other.  (Such objects
appear extended to the eye and are therefore unlikely to be stars.)  These
cuts removed roughly half of the original sources, leaving a sample of
$\sim 5,500$ candidate stars.  The instrumental magnitudes of these
stars were then scaled to $0 \farcs 5$ sized-apertures using
photometry of isolated field stars, and extrapolated to infinite
apertures and placed on the VEGAMAG system using the corrections and
zero points of \citet{sirianni2005}.

We determined the photometric errors and completeness of this catalog
via a series of artificial star experiments.  We used the
frames' PSFs to add 500 stars to each quadrant of the F606W and F814W
images, and repeated our measurement technique, including all the
procedures for the removal of contaminants.  The results were then
compared with the input photometry to determine the completeness 
and accuracy of our measurements.  The process was repeated 100 times 
to produce a total of 200,000 artificial stars with magnitudes
between $30.0 < m_{\rm F814W} < 25.3$ and $-0.7 < (F606W-F814W) < 3.3$.

\subsection{Background Contamination and Aperture Photometry} 

\subsubsection{SExtractor}

In addition to the PSF photometry, we used the SExtractor software
package version 2.3b2 \citep{bertin1996} to detect, measure, and most
importantly, classify objects on our frames.  As in our DAOPHOT
analysis, we ran SExtractor independently on our F814W and F606W
frames, using a $3 \sigma$ detection threshold over a two-pixel
(radius) region, and applied the \citet{sirianni2005} calibrations to
transform our instrumental magnitudes into the VEGAMAG system.  This
yielded a catalog of $\sim 10,100$ objects, with $\sim 5,800$
classified as stars and $\sim 4,300$ classified as galaxies.  The same
artificial stars were then run through the SExtractor routine to
determine the photometric errors and completeness of this procedure.

\begin{figure}
\centerline{\psfig{file=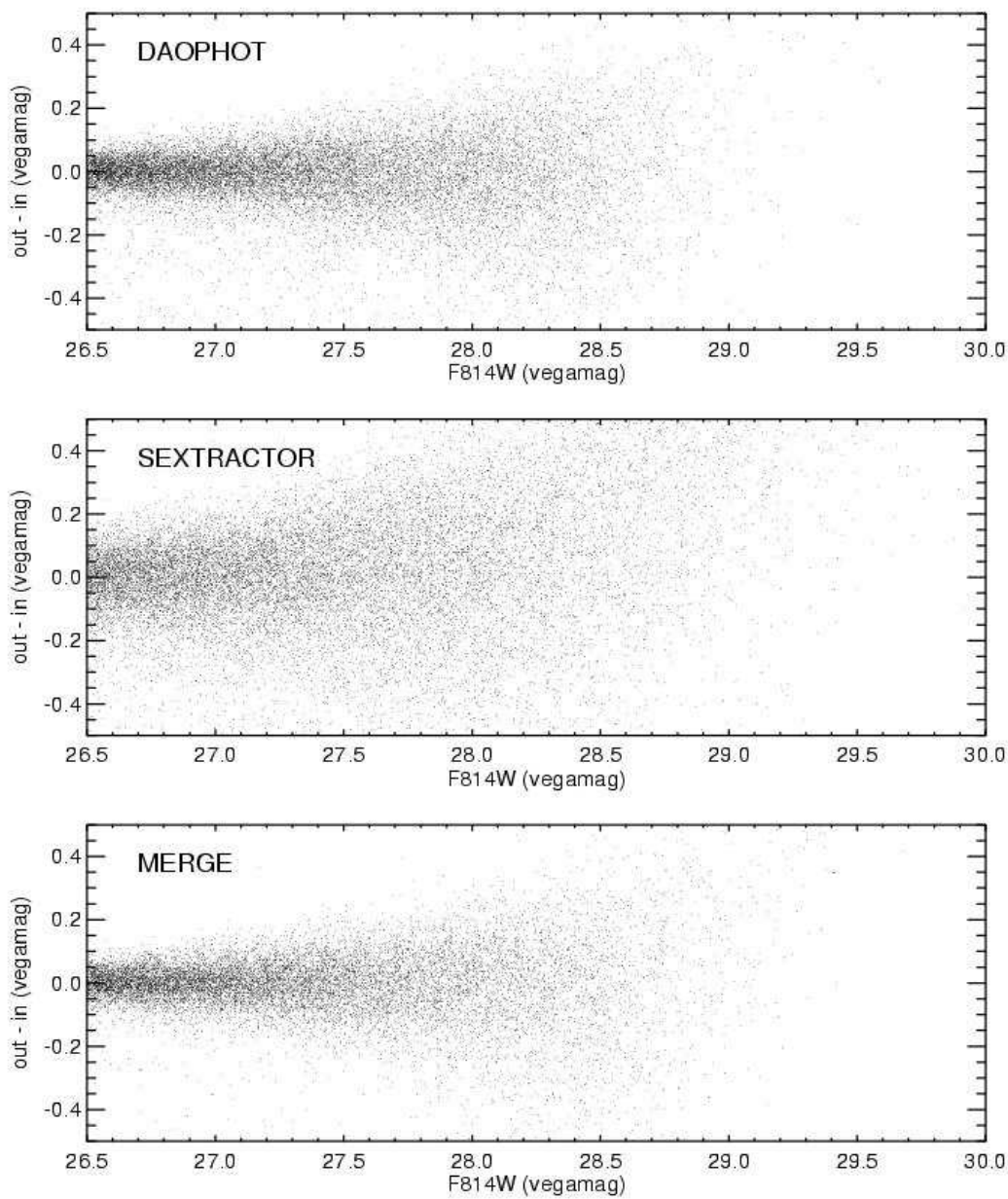,width=5.5in,angle=0}}
\caption{The results of artificial star experiments in the final F814W
frame performed using both DAOPHOT and SExtractor.  {\it Top:} The
measured F814W Vegamag magnitude plotted against the residual
magnitude (measured minus input) as determined by DAOPHOT.  {\it
Middle:} The same diagram using the measurements of SExtractor.  Note
the presence of a systematic error at magnitudes fainter than
$m_{\rm F814W} \sim 28$.  {\it Bottom:} The same diagram using our
combined DAOPHOT and SExtractor selection and measurement criteria.
This procedure provided the best combination of depth and
photometric accuracy.}
\label{comparison}
\end{figure}

A comparison between our DAOPHOT and SExtractor measurements
demonstrates that there is good agreement between the photometric
packages for all but the faintest objects.  Although SExtractor is
able to measure magnitudes for more point sources at the frame limit
than ALLSTAR, the photometric errors associated with SExtractor are
always larger than those found by DAOPHOT and, more importantly, the
SExtractor errors are systematic at the faintest magnitudes.  This is
demonstrated in Figure~\ref{comparison}, which plots the results of
artificial star experiments for both algorithms.  At $m_{\rm F814W} =
27$, the scatter in the SExtractor error is $\sigma \sim 0.12$,
compared to $\sim 0.06$ for DAOPHOT, and by $m_{\rm F814W} = 28.5$,
there is a significant departure from a mean of zero.

\begin{figure}
\centerline{\psfig{file=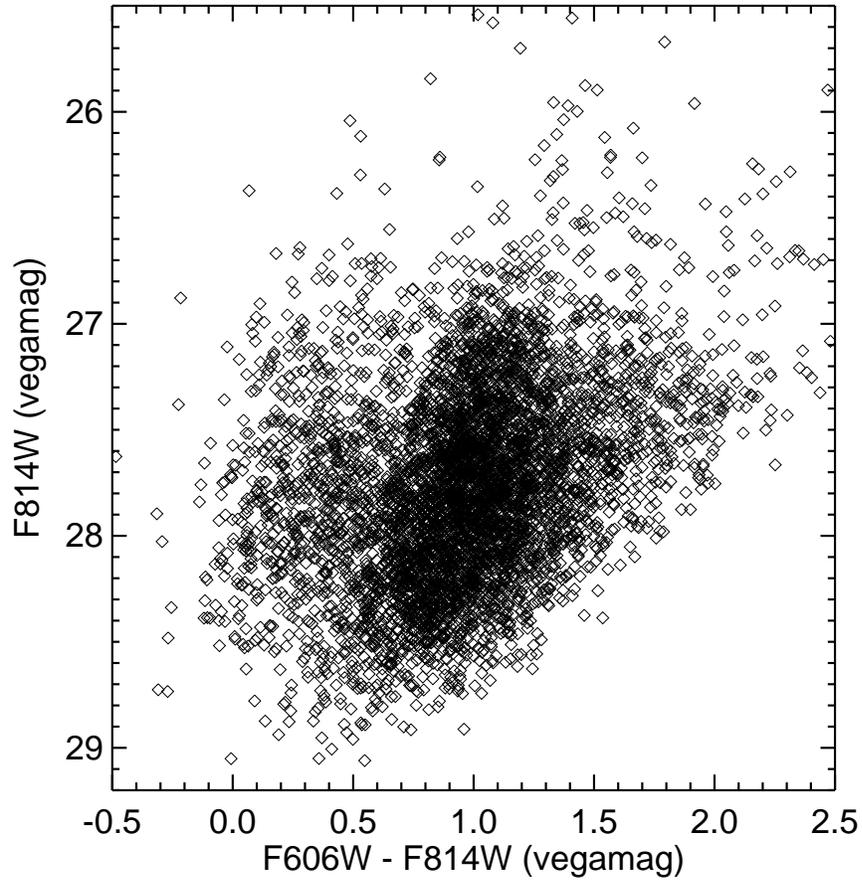,height=5.0in,angle=0}}
\caption{The color-magnitude diagram of objects in our field identified 
by DAOFIND\null.  Only objects that appear in both F814W and F606W and
are classified by DAOPHOT as point sources are plotted.  Note that in 
addition to a red giant branch, there is a substantial number of ``blue''
objects with F606W$-$F814W $< 0.5$.}
\label{rawcmd}
\end{figure}

In contrast to its photometric limitations, SExtractor is much better
than DAOPHOT at discriminating background galaxies from Virgo's
intracluster stars.  This is demonstrated in
Figures~\ref{rawcmd} and \ref{galaxies}, which show DAOPHOT's stellar
CMD and SExtractor's color-magnitude diagrams for stars and galaxies.
From the figures, it is clear that the stellar sample defined by
SExtractor contains far fewer blue contaminants than the DAOPHOT sample.
Specifically, SExtractor's point-source CMD shows only a handful of objects
blueward of the red giant branch,  none of which lie on any stellar
sequence.  Moreover, several of these blue objects appear slightly extended 
to the eye, suggesting that they are marginally resolved background galaxies. 
Conversely, SExtractor's galaxy CMD shows no evidence of a red locus, but does
show a significant population of objects in the blue part of the
diagram.  This strongly suggests that the blue objects present
in DAOPHOT's CMD are not stars.

The results of the above experiment led us to use both SExtractor and
DAOPHOT for our photometry.  In general, we used the SExtractor
classification algorithm to define our sample of intracluster stars, but
then turned to DAOPHOT for the actual photometric measurements.  
The resulting CMD, which includes $\sim 5,800$ stars, is shown in the 
left panel of Figure~\ref{merge}.  It contains the following
objects: 

1) Sources detected in both bands by both programs and classified
as stars by SExtractor.  These objects were assigned their DAOPHOT
magnitudes.

2) Small sources (roundness $< 0.1$) detected in both bands by
DAOPHOT, but not detected in both bands by SExtractor.  These were
also assigned their DAOPHOT magnitudes.

3) Sources detected in only one band by DAOPHOT, but found in
both bands and classified as stars by SExtractor.  These objects were
assigned their DAOPHOT magnitude in one band, and their SExtractor magnitude
in the other.

To determine our photometric error and completeness, an identical set
of criteria were applied to our artificial star experiments.  The
results of these experiments are shown in Figure~\ref{comparison}.
The simulations show that our completeness fraction falls below 50\%
at $m_{\rm F814W} \sim 28.5$ and $m_{\rm F606W} \sim 29.2$.

\subsubsection{Control Field Imaging}

\begin{figure}
\centerline{\psfig{file=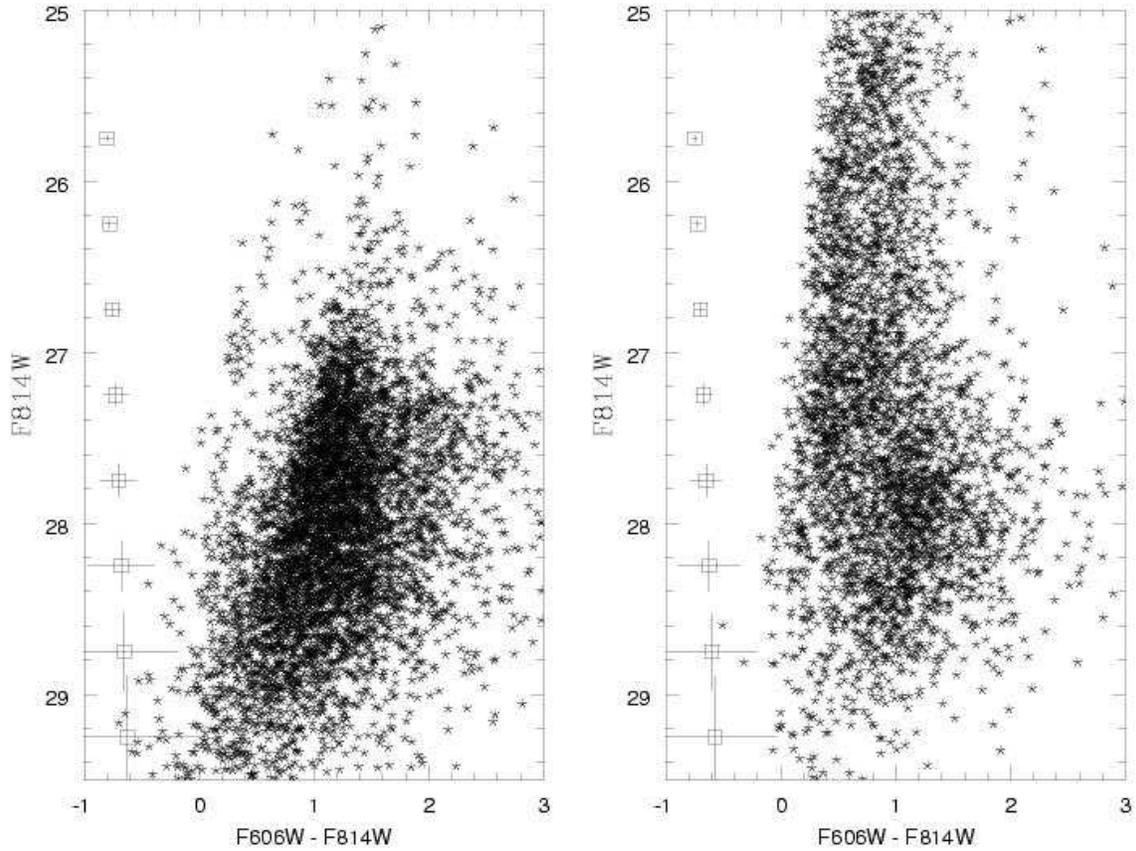,width=6.5in,angle=270}}
\caption{The color magnitude diagram of objects in our field as measured 
with aperture photometry via SExtractor.  Open boxes show the rms errors as a
function of magnitude.  {\it Left:} The objects classified as stars by
the SExtractor star/galaxy value.  The RGB is clearly visible.  {\it
Right:} The objects classified as galaxies by the SExtractor
star/galaxy value.  No obvious RGB is detected.  Note that most
galaxies lie to the blue of the RGB stellar locus.}
\label{galaxies}
\end{figure}

As Figure~\ref{galaxies} shows, even SExtractor mis-classifies some
objects, especially near the frame limit \citep{bertin1996}.
Consequently, in order to further reduce our background contamination,
we used the identical analysis techniques on a different high-Galactic
latitude field with presumably no intracluster stars: the {\sl HST\/}
Ultra-Deep Field \citep[UDF][]{beckwith2006}.  We began by defining
subsets of UDF images which matched the depth and noise
characteristics of our Virgo images.  For the F606W filter, we
collected a series of exposures which matched our 63,440~s image.  To
mimic our F814W dataset, we summed 43,200~s of UDF exposures through
the F775W filter; this longer exposure time compensated for the
filter's 39\% narrower bandpass.

We ran the UDF data through the same analysis routines ({\tt
multidrizzle}, DAOPHOT, SExtractor) as our Virgo data, and converted
the UDF's F775W Vega magnitudes to $I$-band (F814W equivalent)
magnitudes using the transformation coefficients of
\citet{sirianni2005}.  Only 530 sources in the UDF field passed our
point source classification criteria.  Since the UDF presumably
contains no stars at the distance of the Virgo Cluster, this result
implies that only $\sim 10\%$ of the point sources in our program
field are contaminants.  We do note that the UDF is at a lower
Galactic latitude ($b = -54^\circ$) than Virgo ($b = +74^\circ$), thus
our statistical subtraction of point sources may contain an additional
systematic error.  However, simple Galactic models, such as that by
\citet{bahcall1980}, predict that only $\sim 1/4$ of the unresolved
sources in the UDF field belong to the Milky Way galaxy.  Moreover,
the lower Galactic latitude of the UDF is more than offset by its $l =
280^\circ$ longitude, which is more towards the Galactic anti-center
than our Virgo field ($l = 223^\circ$).  Consequently, the Galactic
contribution to the two fields should differ by no more than $\sim
10\%$.  Since there are only $\sim 530$ point sources in the UDF to
begin with, this translates into a $\sim 1\%$ systematic uncertainty
in the Virgo field source counts.

A CMD of the UDF's ``point sources'' is shown in the middle panel of
Figure~\ref{merge}.  Clearly, this diagram shows no evidence of a red
giant branch.  Furthermore, the colors and magnitudes of the
unresolved sources in the UDF field are similar to those of the
non-stellar objects seen in Virgo (\ie\ Figure~\ref{galaxies}).  This
agreement, which is consistent with the results of SExtractor,
confirms that the blue objects in our intracluster star CMD are mostly
marginally resolved or unresolved background galaxies.  It also
validates the use of the UDF data as a control for determining the
amount of contaminants in our intracluster field.

To statistically remove these objects, we deleted from the Virgo point
source catalog the closest match to each UDF point source in
color-magnitude space.  If no Virgo source was within a 0.15~mag
radius circle of the UDF source, no source was removed.  This radius
was chosen to be similar to the photometric errors of faint sources
and to the separation between points in color-magnitude space for
bright objects.  However, the exact radius has little impact on the
results of the subtraction. If a 0.2~mag radius is adopted, only 12
more stars are removed from the diagram.  If this restriction is
removed entirely, then the background-subtracted CMD becomes somewhat
less populated, though the metallicity and age distributions inferred
for the population remain essentially unchanged (see below).  As
implemented, the procedure resulted in the exclusion of 470 (out of a
possible 530) objects from the point source catalog, and reduced the
expected amount of contamination in our intracluster field to $\sim
1\%$.  The 5,300 stars remaining in Virgo intracluster sample are
shown in the right panel of Figure~\ref{merge}.

\subsection{Measuring the Metallicity Distribution}\label{starfish}

\begin{landscape}

\begin{figure}
\centerline{\psfig{file=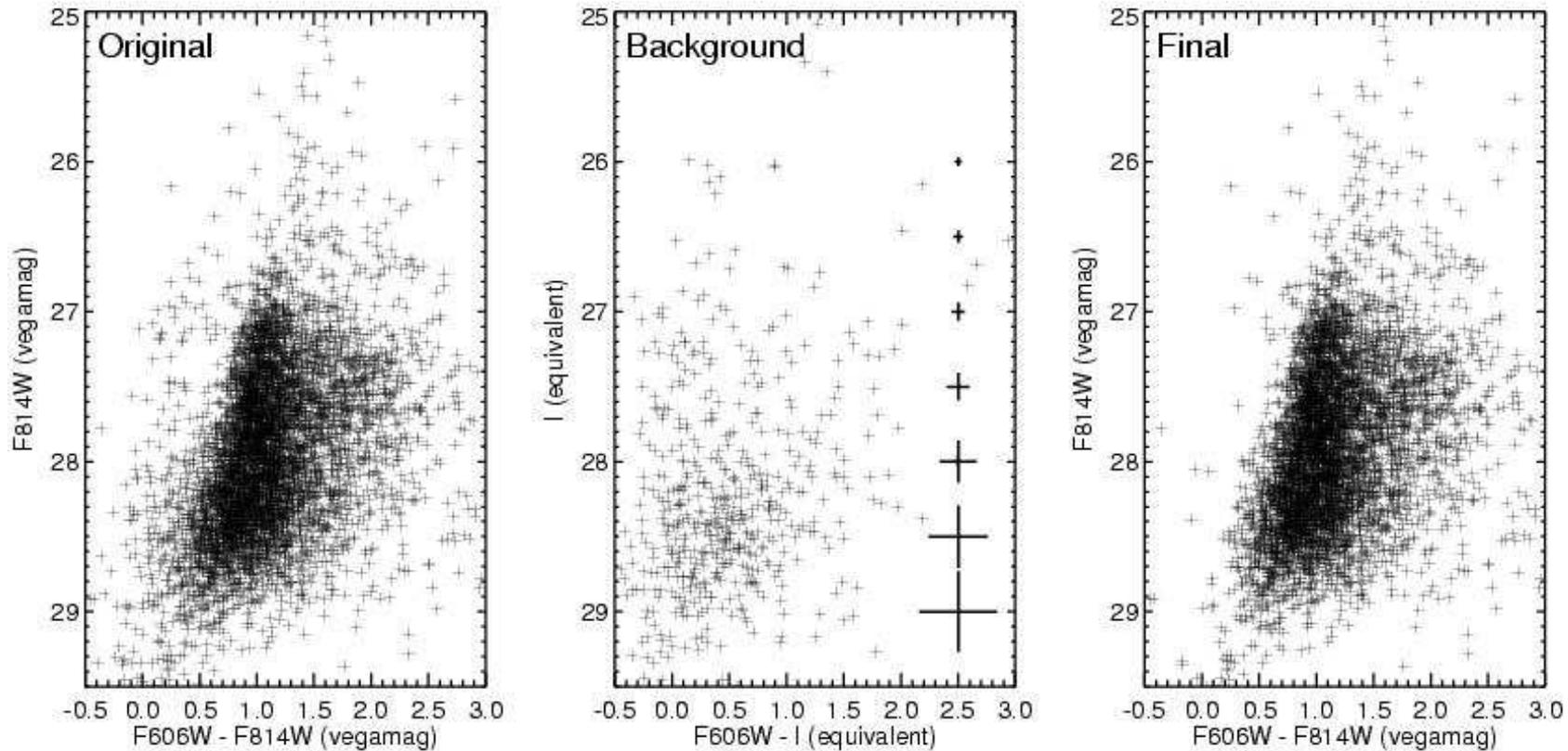,height=4.5in,angle=0}}
\caption{{\it Left:} The color magnitude diagram of objects in our
field that match the combination of DAOPHOT and SExtractor criteria for
stars.  No background objects have been removed.  {\it Middle:} The
CMD of objects in our UDF control field that match the combination
of DAOPHOT and SExtractor criteria for stars. Bold error bars show the
rms errors for all of the data (see Table~\ref{errors}) as a function
of magnitude.  {\it Right:} The final CMD used for measuring the
metallicity distribution of Virgo's intracluster component, where
background objects have been subtracted away.}
\label{merge}
\end{figure}

\end{landscape}

A quick look at the final CMD of Figure~\ref{merge} demonstrates that
the width of Virgo's red giant branch is significantly greater than
the scatter due to photometric errors (given in Table~\ref{errors}).
This demonstrates that the intracluster stars possess a range of
metallicities \citep[see arguments in, \eg][]{zinn1980}.  Also obvious
is the small number of stars brighter than the RGB tip, which is
located at $m_{\rm F814W} \sim 27$.  This small quantity is strongly
suggestive of a very old population, with $t \sim 10$~Gyr.  In
addition, the CMD contains few, if any, objects with the colors of
upper-main sequence or blue-loop stars $(V-I < 0$).  This excludes the
possible presence of a very young component ($< 30$~Myr) in this
region of intracluster space.  Finally, the RGB tip appears sharp,
suggesting that the stars in our field are not distributed over a very
wide range in distance.

To quantify these conclusions, we analyzed the CMD of
Figure~\ref{merge} with the StarFISH $\chi^2$ minimization code of
\citet{harris2001}.  This software package treats an observed
color-magnitude diagram as a superposition of eigenpopulations, each
defined through the convolution of a theoretical isochrone with the
photometric error and completeness function.  The code then finds the
linear combination of parameters which minimizes the $\chi^2$
difference between the observed Hess diagram and the synthetic
diagram.  The base isochrones chosen for analysis were those of
\citet{girardi2002}, which have been updated to reproduce colors from
the {\it HST ACS\/} filter
set\footnote{http://pleiadi.pd.astro.it/isoc\_photsys.02/isoc\_photsys.02.html}.
This dataset includes models with $\log$ ages of 8.5, 9.0, 9.35, 9.65,
9.9, and 10.1, and metallicities of $Z = 0.0001, 0.0004, 0.001,
0.004, 0.008$ and 0.019.  Because our observations only extend
$\sim 1.5$~mag down the red giant branch (with $>$50\% completeness),
our analysis was completely insensitive to the assumed slope of the
initial mass function.  (We adopted a \citet{salpeter1955} slope for
this relation.)  The photometric error and completeness estimates were
taken directly from the results of our artificial star experiments, and
the fits were performed using both 0.1~mag and 0.2~mag bins sizes to check 
for consistency.  Finally, to refine our sensitivity to metallicity, we
also included in our input grid an interpolated set of isochrones at 
$Z = 0.0025$.  We experimented with using additional isochrone grids in the
metallicity range between the $Z = 0.004$ and $Z = 0.019$, but these did
not significantly alter our results and added many more degrees of
freedom to the model.  

The only remaining parameters required for the model were the stars' 
reddening and distance.  For the foreground reddening, we adopted the 
COBE/DIRBE-based differential extinction of $E(B-V) = 0.025$ 
\citep{schlegel1998}.  This value is close to the \citet{burstein1982} 
estimate of $E(B-V) = 0.030$ and, with a \citet{cardelli1989} reddening law, 
translates into total extinctions of $A_{\rm F606W} = 0.069$ and 
$A_{\rm F814W} = 0.045$ \citep[see][]{sirianni2005}.  

Unfortunately, obtaining a distance constraint is not so simple.
Thanks in large part to the {\sl HST Key Project\/}
\citep{freedman2001}, there is now general agreement about the
distance to the main body of the Virgo Cluster, with Cepheids
\citep{freedman2001}, the surface brightness fluctuation method
\citep[][\ as recalibrated by \citet{jensen2003}]{tonry2001} and the
planetary nebula luminosity function \citep{jacoby1990, ciardullo1998}
all producing distances between 14.5 and 15.8~Mpc.  However, the
situation with regard to our intracluster stars is more complex.  The
Virgo Cluster is highly elongated along our line-of-sight, extending
from $\sim 14$~Mpc to $\sim 22$~Mpc \citep{west2000, solanes2002}, and
the complex intracluster features seen by \citet{mihos2005} likely
span a range of distances.  In addition, the intracluster stars
plotted in Figure~\ref{merge} are not necessarily associated with the
core of Virgo.  As illustrated in Figure~\ref{mihos}, our field is
located midway between Virgo's central cD galaxy M87, and M86, a
galaxy whose group is known to be falling into Virgo from behind
\citep[\eg][]{jacoby1990, binggeli1993, bohringer1994}.  Thus, the
intracluster stars we observe may be from either (or neither) of these
systems.

Fortunately, the RGB is a distance indicator in itself: for old
metal-poor populations, the absolute magnitude of the RGB brightest
stars is independent of metallicity.  Moreover, though the RGB tip
does fade in metal-rich systems, the shape of the giant branch and its
maximum magnitude are still predictable.  \citep{dacosta1990,
lee1993}.  Consequently, we can use the StarFISH program to find the
distance that best fits the RGB of our field.  By doing so, we can
reduce the effects of any distance-dependent systematic error, and
derive our metallicities in a completely self-consistent manner.

To do this, we ran the StarFISH code using a series of assumed
distances, each time binning the synthetic CMD into 0.1~mag $\times$
0.1~mag intervals, ignoring the asymptotic giant branch (AGB)
contribution to the CMD for the oldest ages, and using its downhill
simplex ``amoeba'' algorithm \citep{press1992} to minimize the
difference between the model and the observations.  The uncertainties
in this model were then defined by starting from the best fit solution
and varying each parameter in turn, until the overall $\chi^2$
increased by 5\%.  Therefore the errors do not take into account any
problems that may exist in the RGB models themselves. The procedure
yielded a best-fitting distance modulus of $(m-M)_0 = 31.05 \pm 0.05$
(16 Mpc).  Our subsequent metallicity and age measurements for the
stellar population depend little on this exact number.

As mentioned above, in performing our Hess diagram fits, we ignored
the contribution of AGB stars in our oldest models.  Our field
contains very few stars brighter than the RGB, thus suggesting that
the intracluster stars of Virgo are mostly very old ($ > 10$~Gyr).
Yet even our oldest \citet{girardi2002} models over-predict the number
of stars in this part of the color-magnitude diagram.  This is a
general property of all our fits: regardless of age, metallicity, or
initial mass function, the AGB population produced by the models is
much greater than that seen in the data.  Consequently, the presence
of this component limited the quality of our fits to $\chi^2/\nu =
3.8$.

One possible explanation for the AGB anomaly lies in this isochrones'
treatment of mass-loss.  As a population ages and its turn-off mass
declines, the details of mass-loss become more and more important, and
a slight increase in the mass-loss efficiency, $\eta$, can lead to the
premature termination of the AGB phase, or even its complete
elimination \citep[\eg][]{greggio1990}.  Moreover, the initial
mass-final mass relation, which is used to calibrate the
\citet{girardi2002} AGB mass-loss law, is particularly ill-constrained
at the low-mass limit \citep{herwig1995, weidemann2000}.  Thus, it is
not surprising that models have difficulty with this phase of stellar
evolution.  Indeed, an examination of the color-magnitude diagrams of
some of the Galaxy's oldest globular clusters confirms the rarity of
bright AGB stars \citep[\eg\ see][]{buonanno1985, durrell1993}.

Because the AGB phase is difficult to model and likely to become less
populated as the system's age increases, we experimented with varying
the importance of this stage of evolution.  In one case, we removed
the phase entirely from all of our input isochrones; in another, we
eliminated the AGB in all isochrones older than 10~Gyr, but kept the
stage in the younger models. These experiments, described in detail in
\S~3.2, provided valuable insight about the metallicity distribution
of the stars in our field as well as the limitations of our data.

\section{Results}\label{results}

\subsection{Spatial Distribution}

The Virgo Cluster is known to have a significant line-of-sight depth,
which ranges from $\sim 15$~Mpc to $\sim 22$~Mpc \citep{west2000,
solanes2002}.  Yet, from the paucity of stars dectected above the
nominal tip of our red giant branch, it is clear that the stars of our
intracluster survey region are not nearly so spread out.  This
suggests that Virgo's diffuse population is structured.

\begin{figure}
\centerline{\includegraphics[width=5.5in,angle=0]{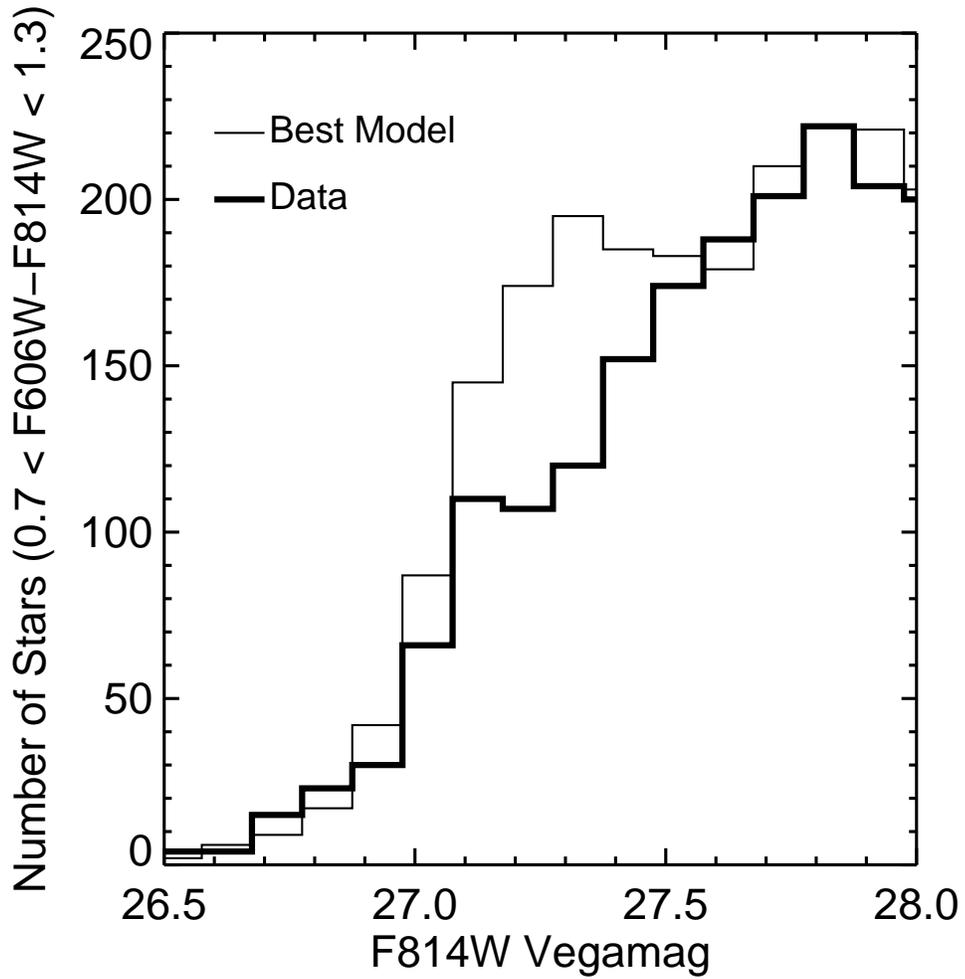}}
\caption{Histograms of the tip of the red giant branch in our best
model CMD (light histogram) and our data (dark histogram).  While both
show a sharp edge at F814W$\sim$27, the edge seen in the data is
shorter and shows a level portion before recovering to the level of
the model, consistent with the possibility of a few Mpc of depth in
our field along the line of sight.}
\label{trgb}
\end{figure}

To quantify this statement, we can examine the luminosity function of
likely RGB stars in Virgo's intracluster space.  This is done in
Figure~\ref{trgb}, where we compare the observed distribution of stars
to that predicted by our best-fitting model.  Both the model and the
data display a sharp edge at F814W $\sim 27$, which is indicative of
the tip of the red giant branch.  However, the discontinuity present
in the data is much less pronounced than that in the model.  This
shallower break, along with the recovery at F814W $\sim 27.4$
indicates that at least some of our RGB stars are distributed along
the line-of-sight.

The strength of the break in the Virgo luminosity function
demonstrates that $\sim 70\%$ of the stars in our intracluster field
are at a common distance of $\sim 16$~Mpc ($\pm 0.1$~mag).  The other
$\sim 30\%$ of the population extends beyond the main clump by $\sim
0.3$~mag.  Interestingly, this spread is consistent with the
line-of-sight depth inferred from intracluster planetary nebula
photometry, though the PN data are more sensitive to objects in the
foreground, rather than background \citep{ciardullo1998,
arnaboldi2002, feldmeier2004}.  The spread in distance is also
consistent with conclusions of \citet{arnaboldi2004}, who used radial
velocity measurements to show that Virgo's intracluster planetaries
are kinematically not well-mixed.

If the intracluster stars of Virgo are not well-mixed, then our {\sl
ACS\/} field may not be typical of the rest of Virgo, and the
metallicity distribution that we determine may not be representative
of that of the Virgo cluster as a whole.  Recent cosmological
simulations \citep{murante2006} predict that, though most intracluster
stars have their origins in massive galaxies, a sizeable fraction of
the population comes from the dissolved remains of lower-mass objects.
These models also predict that the ratio of these two components will
change with distance: the larger the distance from the
cluster-dominant galaxy, the higher the fraction of disintegrated
galaxies.  If true, then the metallicity distribution function of the
diffuse component should also change, and the values measured in our
{\sl ACS\/} field, which is $\sim 200$~kpc from M87, might be lower
than the average for the cluster.

While large-scale variations in the intracluster light metallicity can
only be checked by analyzing further judiciously spaced fields, we can
use our current data to test for small-scale variations in the
intracluster stellar population.  To do this, we created two samples
of intracluster stars from our final catalog, the first consisting of
stars brighter than $m_{\rm F814W} = 28$, and with $0.65 < {\rm
F606W}-{\rm F814W} < 1.0$ (the color of the tip of the RGB for [M/H]
$< -1.7$), and the other comprised of stars brighter $m_{\rm F814W} =
28$ and with $1.25 < {\rm F606W}-{\rm F814W} < 2.4$ (the color of the
RGB tip for [M/H] $> -1.2$).  This procedure produced two sets of 1600
stars; we then analyzed the spatial distribution of these stars by
dividing our {\sl ACS\/} field into a $6\times 6$ grid, (ignoring the
region surrounding the resolved dwarf galaxy and the chip gap).  This
grid size provided enough resolution to detect spatial variations
within our field while limiting the errors due to counting statistics
to less than 20\%.

Overall the distribution of intracluster stars in our field is fairly 
homogeneous.   However, the bluest (most metal-poor) stars show significant 
inhomogeneities on the scale of about half the field size 
($\sim 5~{\rm kpc}$).  In particular, the dwarf galaxy described in 
\citet{VICS3} is part of a higher density region of metal-poor
stars, which is possibly a stream that runs through the field.   When
compared to a random distribution, the observed distribution of metal-poor
stars has a $\chi^2/\nu = 7$, which equates to a probability less than
$10^{-30}$.  For comparison, similar analyses for the metal-rich stars
and the full sample yield $\chi^2/\nu = 2.7$ ($P \sim 4.7 \times 10^{-7}$)
and $\chi^2/\nu = 2.6$ ($P \sim 1.4 \times 10^{-6}$).   The fact that the
metal-poor stars show considerably more structure than the metal-rich
objects presumably means that these objects either were liberated more
recently or originated in a smaller object.  A detailed discussion of
the spatial distribution of these intracluster giants will be presented
in a separate paper.

\subsection{Age and Metallicity Distribution}

\subsubsection{Fitting the CMD}

\begin{figure}
\centerline{\psfig{file=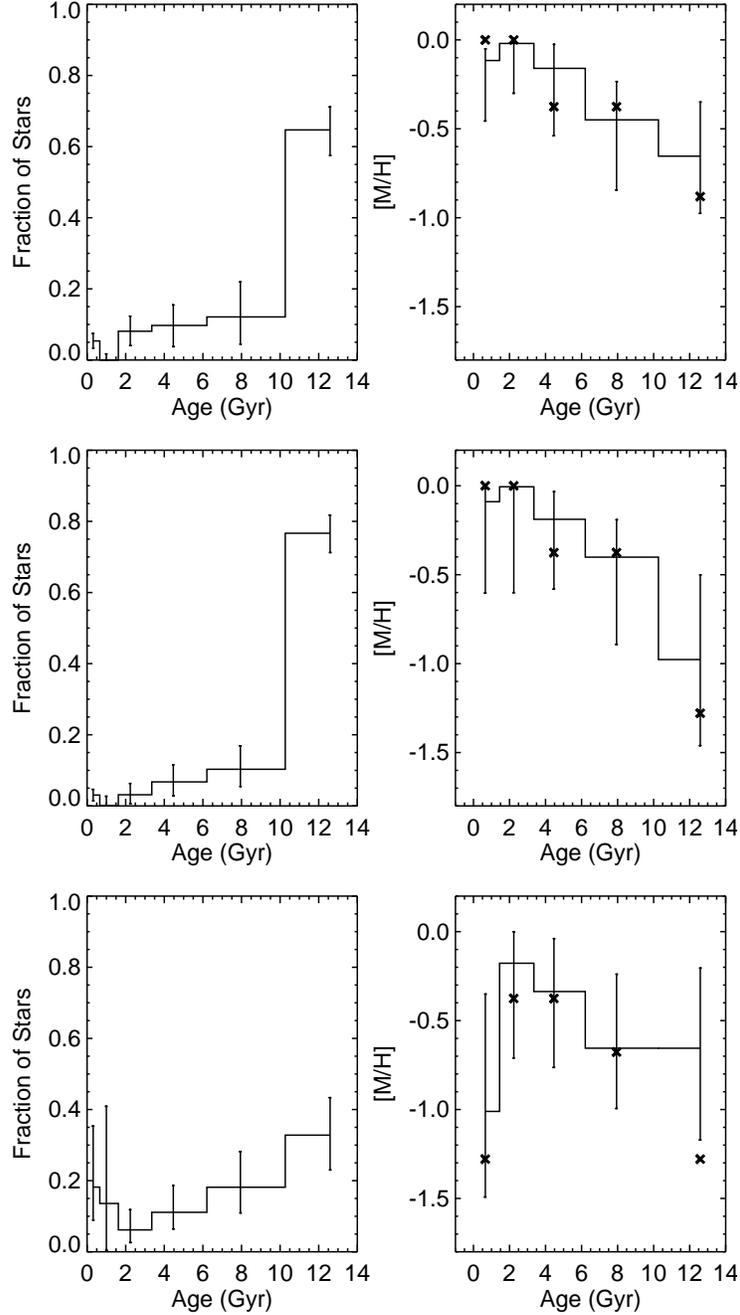,height=7.0in,angle=0}}
\caption{{\it Top row: }The age distribution of Virgo's intracluster
stars as derived by the StarFISH minimization code from the raw
Girardi et al.~isochrones.  {\it Top Left:} The fraction of stars of
each age present in the model.  {\it Top Right:} A histogram and error
bars marking the mean metallicity and uncertainty of each age bin.
The dark X's mark the median metallicity of the bin. {\it Middle row:
} Same as {\it Top} but using the Girardi et al.~isochrones with the
AGB contribution removed from models older than 10~Gyr. {\it Bottom
row: } Same as {\it Top} but using the Girardi et al.~isochrones with
the AGB contribution completely removed from the models.}
\label{ad}
\end{figure}

\begin{landscape}

\begin{figure}
\centerline{\psfig{file=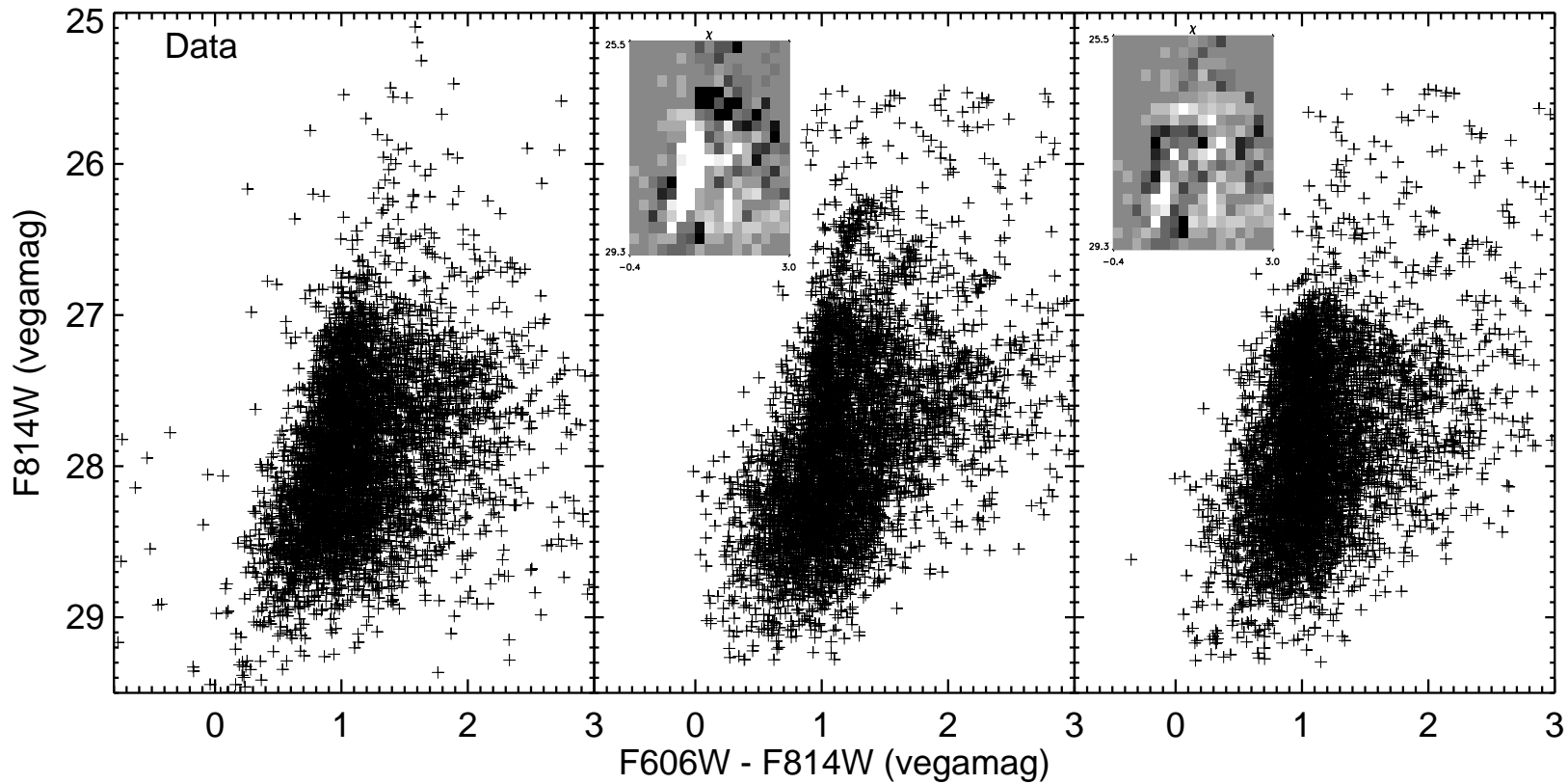,height=4.5in,angle=0}}
\caption{{\it Left:} The observed CMD of the intracluster stars in our field.
{\it Middle:} Our best-fit model CMD with the AGB included in the stellar 
evolution models. The inset grayscale plot shows the $\chi^2$ distribution 
where darker areas represent regions where the predicted stellar density 
is larger than that observed.  Notice the under-reproduced
metal-poor RGB and the over-reproduced metal-poor AGB\null. {\it Right:} Our
best-fit model CMD with the AGB excluded from stellar evolution
models with $t \geq 10$~Gyr. The inset grayscale plot again shows the
$\chi^2$ distribution where darker areas represent regions where the
model over-predicts the number of stars.  Note the excellent agreement
between the data and the model.}
\label{repop}
\end{figure}

\end{landscape}

\begin{figure}
\centerline{\psfig{file=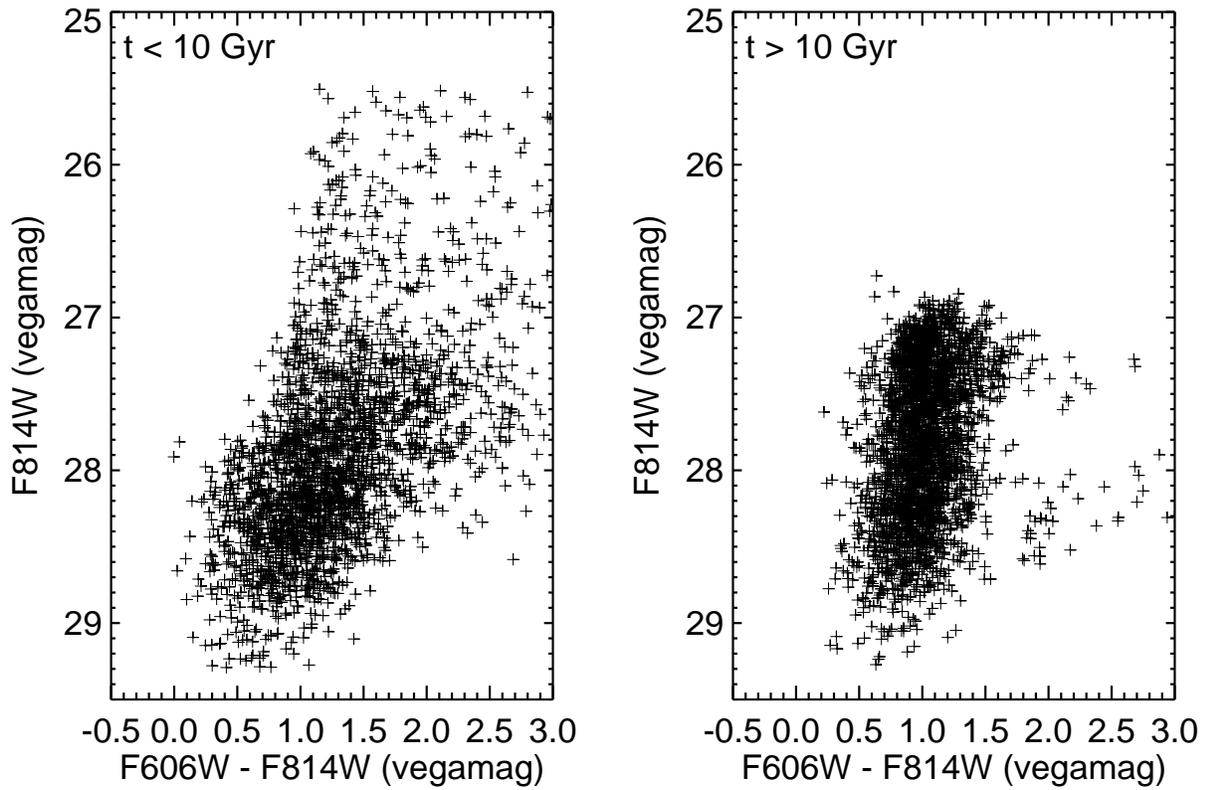,width=6.5in,angle=0}}
\caption{{\it Left:} The contribution of stars younger than 10~Gyr to
our best-fit model color-magnitude diagram.  {\it Right:} The
contribution of stars older than 10 Gyr to our best-fit model
color-magnitude diagram; the AGB was excluded from the isochrones for
these ages.}
\label{oy}
\end{figure}

The results of the StarFISH analysis are shown in
Figures~\ref{md}--\ref{oy}.  Figures~\ref{md} and \ref{ad} show the
best-fitting metallicity and age distributions; Figure~\ref{repop}
compares our model CMDs with the actual data.  Note that because our
observations do not reach the red clump, the horizontal branch, or the
main-sequence turn-off, our ability to determine the age distribution
of Virgo's intracluster stars is limited.  Younger stars do populate
the RGB somewhat differently than older objects --- in general, they
are slightly bluer for the same metallicity --- but most of our
leverage on age comes from the AGB.  Unfortunately, as discussed
above, models of AGB evolution are fraught with uncertainty, and any
result which depends solely on the fit to this area of the
color-magnitude diagram is not robust.  Moreover, all models which use
the \citet{girardi2002} isochrones, regardless of age, metallicity, or
initial-mass function, overpredict the number of Virgo AGB stars.
Thus, to examine the effect of the AGB on our analysis, we created
three different sets of models: one based on the raw isochrones with
the AGB included, one in which the AGB component was excluded from the
very oldest ($t > 10$~Gyr) models, and one in which no AGB stars were
fit.

The quality of these fits varied considerably.  When the AGB is
included in the fits, $\chi^2/\nu = 3.8$; when this component is
removed (and the fit is restricted to stars with $m_{\rm F814W} >
26.5$), $\chi^2/\nu$ drops to = 1.9.  The best results are obtained
from the model which includes the AGB for all but the oldest stars; in
this case $\chi^2/\nu = 1.6$.  Figure~\ref {repop} illustrates the
reason for this improvement.

As Figure~\ref{repop} illustrates, when we fit the observed CMD using
the raw \citet{girardi2002} isochrones, our best-fitting model
underpredicts the number of metal-poor RGB stars ($m_{\rm F814W}
> 27$, F606W$-$F814W$ < 1.0$), but overpredicts the number of old,
metal-poor AGB stars ($m_{\rm F814W} < 27$, F606W$-$F814W$ < 1.0$).
Clearly, metal-poor RGB stars are present, but the model cannot match
the ratio of RGB stars to AGB stars, even for the oldest ages.  When
the AGB is removed from the oldest isochrones, the fit is much
improved: StarFISH can increase the number of old, metal-poor ([M/H]
$< -1.5$) stars to match the RGB, without overpopulating the AGB
component.  If the AGB is removed completely from the analysis and the
fit is restricted to the RGB stars, then a full magnitude of CMD space
is lost, as magnitudes brighter than $m_{\rm F814W} < 26.5$ are not
included in the fit.  The smaller number of degrees of freedom causes
a slight increase in $\chi^2/\nu$, but the fit still requires a
metal-poor ([M/H] $< -1.5$) component of the population.

Finally, we note that, although we have not included it in our
$\chi^2$ calculations, there is one additional constraint that can be
applied to our analysis.  According to the ultra-deep surface
photometry of \citet {mihos2005}, the total $V$-band surface
brightness in our field is $\mu_V \sim 27.7$.  About half of this is
due to background galaxies: if we form a $V$-band $\log N - \log S$
curve from the non-point sources in our field \citep[using the
transformations of ][]{sirianni2005} and extrapolate the observed
power law to $m_V \sim 30$, then the total surface brightness in
galaxies is $\mu_V \sim 28.6$.  This implies that the total {\it
stellar\/} surface brightness in our field is $\mu_V \sim 28.3$.

We can compare this value to the results of the StarFISH code.  Our
fits produce models for the age and metallicity of Virgo's
intracluster RGB and AGB populations.  If we use a \citet{salpeter1955}
initial mass function to extrapolate the star counts down the main
sequence (to $M_{\rm F814W} = +7$), then our fits also yield the total
number of stars in our field and the stellar surface brightness.  This
value must come close to matching the results of the broadband surface
photometry for the model to be plausible.

In fact, the best-fit solution without AGB stars far exceeds the
observational constraint, producing a $V$-band surface brightness of
$\mu_V \sim 27.1$.  The high surface brightness occurs with the
addition of intermediate-aged, metal-poor stars, which require the
presence of bright main sequence turnoff stars below our magnitude
limit. If such bright main-sequence stars were present in our field,
the total surface brightness would clearly be higher (even though we
would not detect the stars in our data).  On the other hand, if the
metal-poor stars are old, the associated main sequence stars below our
mangitude limit are fainter, keeping the total surface brightness
within the observational constraint.

Because the presence of an intermediate-age, metal-poor population
would apparently increase the total surface brightness of the
intracluster medium to a level that is inconsistent with observations,
the solution obtained from completely removing the AGB from the
models, although instructive, is not viable.  The other models both
produce a value for the integrated light that is much closer to that
observed, $\mu_V \sim 28.1$.  Because of the uncertainties associated
with AGB mass-loss (especially in old, metal-poor stars), and because
the solution that omits the AGB for $t > 10$~Gyr populations has a
much better $\chi^2/\nu$ than the models which include the AGB, this
model is our preferred solution.

\subsubsection{Metallicity and Age}

\begin{figure}
\centerline{\psfig{file=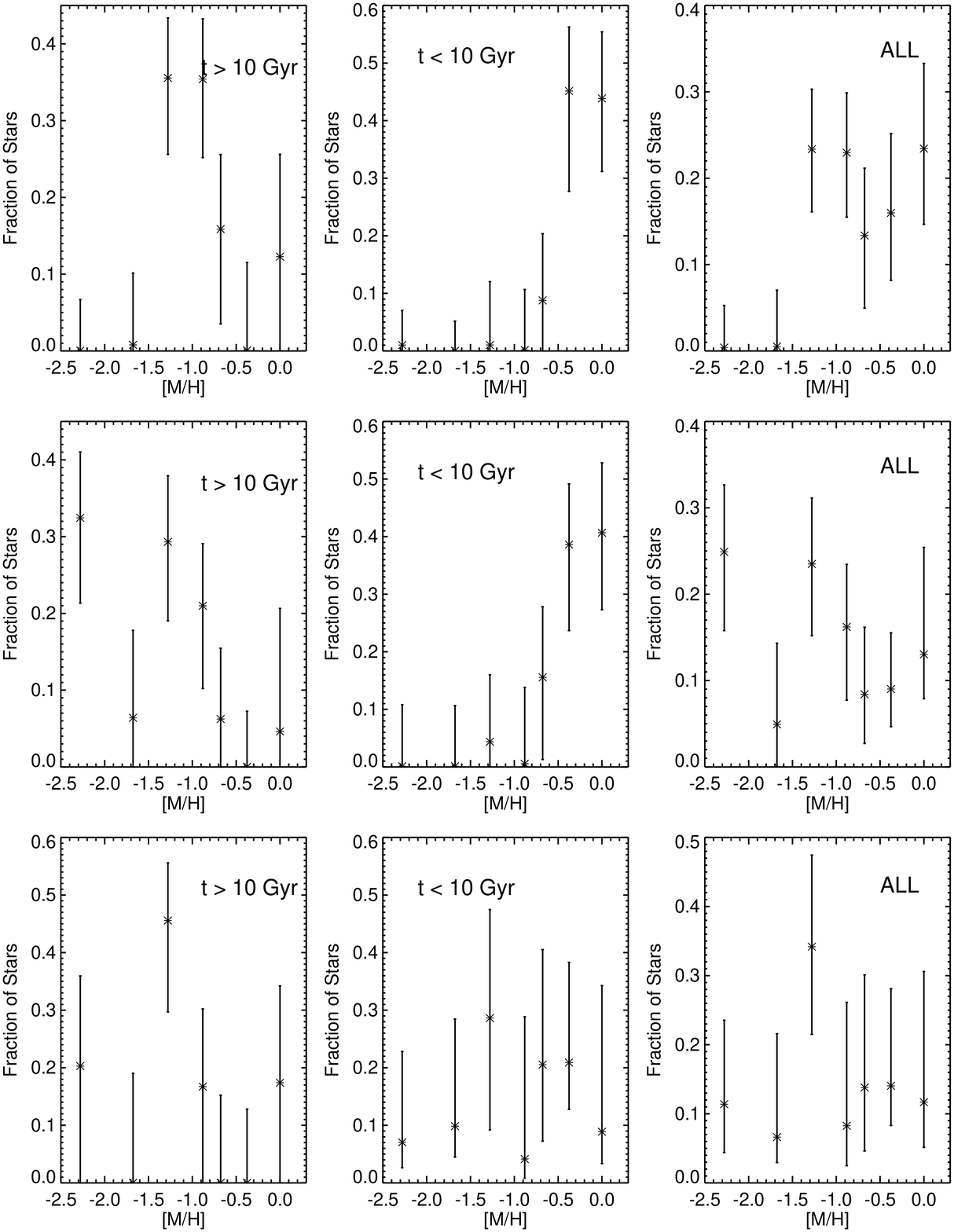,width=5.5in,angle=0}}
\caption{The metallicity distribution function of intracluster stars
inferred from the StarFISH CMD model fits.  {\it Top row:} The
resulting metallicity distribution when the fits are performed using
the raw Girardi isochrones.  {\it Top Left:} The fraction of stars in
each metallicity bin with ages $>$10 Gyr.  {\it Top Middle:} The
fraction of stars in each metallicity bin with ages $<$10 Gyr.  {\it
Top Right:} The overall fraction of stars in each metallicity
bin. {\it Middle row:} Same as {\it Top} but using the Girardi et
al.~isochrones, with the AGB contribution removed from models older
than 10~Gyr. {\it Bottom row:} Same as {\it Top} but using the Girardi
et al.~isochrones, with the AGB contribution completely removed from
the models.}
\label{md}
\end{figure}

In contrast to the problem of age, our F606W$-$F814W color-magnitude
diagram is a good probe of population metallicity.  As Figure~\ref{md}
shows, the metallicity distribution function of Virgo's intracluster
stars is also affected by the exclusion of the AGB from the models,
and the effects help to show the problem with including the AGB for
all ages.  After using our experiments to decipher the most likely age
and metallicity distribution of the stars in our field, we can begin
to compare the results with those of numerical simulations.

As Figure~\ref{ad} illustrates, Virgo's intracluster population is
dominated by old, metal-poor stars.  Our best-fitting model suggests
that 70-80\% of the stars have ages that are greater than $\sim
10$~Gyr; these stars have a median metallicity of [M/H] $\sim$ -1.3, a
mean metallicity of [M/H] $\sim -1.0$ and range from $-2.5\lap {\rm
[M/H]} \lap -0.7$.  However, there is some evidence for the existence
of a younger, metal-rich component.  To reproduce the observed color-
magnitude diagram, and not overpopulate the AGB, at least two age
components are needed.  Moreover, as Figure~\ref{md} demonstrates, the
metallicity distribution function of the younger ($t < 10$~Gyr)
component clearly peaks at a much higher value than that for the
oldest stars.  According to our best-fit model, the younger (more
metal-rich) component makes up $\sim 46\%$ of the stars present in our
HR diagram.  Note, however, that when integrated over the entire mass
function, the importance of this population is much less, \ie\ only
20-30\% of the population.

If StarFISH is forced to fit the AGB population using the raw \citet
{girardi2002} isochrones, then the fraction of metal-poor stars
inferred for Virgo is significantly less, and there are almost no
stars more metal-poor than [M/H] $< -1.5$.  However, the blue side of
the RGB is not properly reproduced, demonstrating that this result is
not reliable. In both cases where the AGB is at least parially
excluded from the isochrones, the results show a very metal-poor
component ([M/H]$\sim$-2), which is clearly seen in the observed RGB.
Our surface brightness tests further suggest that the metal-poor
component is old (see \S~3.2.1 for details).

\begin{figure}
\centerline{\psfig{file=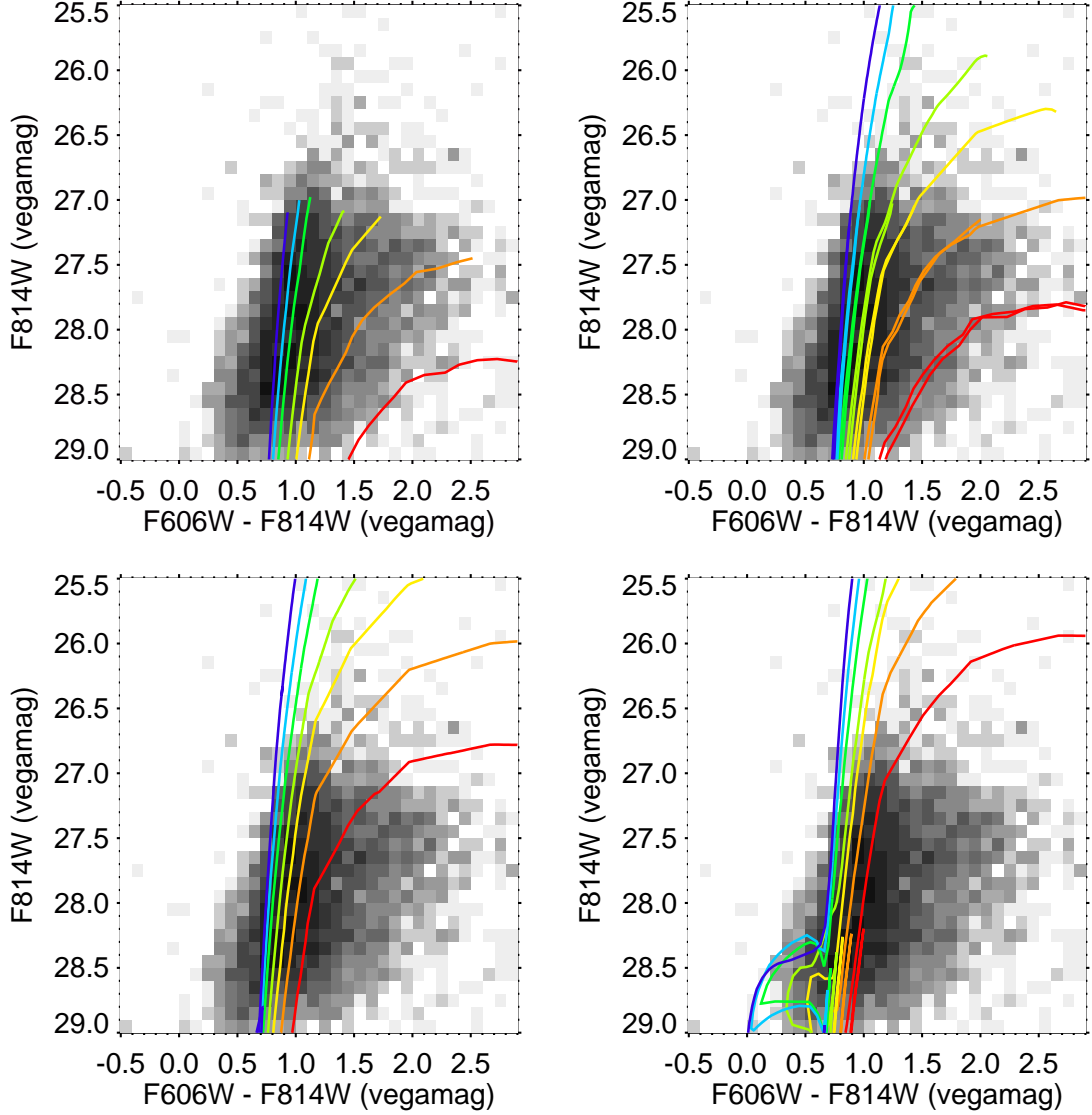,height=6.0in,angle=0}}
\caption{A subset of the Girardi et al.~isochrones for our 
{\sl HST ACS\/} filters superposed on the observed Hess diagram
of our Virgo field.  In each panel, the isochrones represent
stars with metallicities of $Z=0.0001, 0.0004, 0.001, 0.0025,
0.004, 0.008,$ and 0.019, with redder isochrones corresponding to
higher values of $Z$. Each panel displays a different age.
{\it Upper left:} $\log({\rm age}) = 10.1$; the contribution of the AGB was
removed from these isochrones.  {\it Upper right:} $\log({\rm age}) =
9.65$. {\it Lower left:} $\log({\rm age}) = 9.0$. {\it Lower right:}
$\log({\rm age}) = 8.5$.
}
\label{hess}
\end{figure}

Our observations are not directly sensitive to metal-rich stars.
Because of the effects of line blanketing, old solar-metallicity stars
at the tip of red giant branch are not brighter than $m_{\rm F606W}
\sim 30.5$ (see Figure~\ref{hess}).  Since this is more than magnitude
below our detection limit, our analysis does not formally exclude the
existence of such a population.  However, two pieces of evidence argue
against this possibility.  The first comes from the number of
unmatched point sources detected on the F814W image.  In our Virgo
intracluster field, there are 960 point-like sources brighter than
$m_{\rm F814W} = 28.5$ present in F814W but invisible in F606W; in the
UDF, the number of such objects is 12.  Since a total of $\sim 5,300$
stars were detected in both F814W and F606W, this number suggests that
the fraction of very metal-rich stars in our field is $\sim 15\%$.
This is consistent with the StarFISH results.

A second limit on the contribution of old, very metal-rich stars to
the intracluster population comes from the observed background surface
brightness.  As discussed above, after the contribution of background
galaxies is removed, the implied surface brightness of our best-fit
model is an excellent match to the surface photometry results of
\citet{mihos2005}.  This agreement leaves little room for an extra
component of metal-rich stars.

Evidence from ultra-deep surface photometry \citep{mihos2005},
planetary nebula spectroscopy \citep{arnaboldi2004}, and our star
counts all suggest that Virgo's intracluster component is not 
well-mixed.  This opens up the possibility that our results are
not representative of the Virgo Cluster as a whole.  Nevertheless,
we can draw some general conclusions from our observations.
The most obvious is that Virgo's intracluster stars do not come from
a Population~III source.  Since the fraction of extremely metal-poor
stars is low, the stars must have been predominantly formed inside of
galaxies and subsequently removed.  

Finally, the metallicity and age distributions derived by StarFISH are
similar to that generally predicted by simulations \citep{murante2004,
willman2004, sommer2005}.  While the details of these models vary,
most agree that the diffuse component should be old.  The data
displayed in Figure~\ref{ad} support this result.  However, the
metallicities derived for our intracluster field are somewhat lower
than that expected by theory.  For example, \citet{sommer2005} predict
a metallicity distribution function that peaks in the range [Fe/H]
$\sim -0.3$~--~$-0.4$, while \citet{willman2004} infer metallicities
that are similar to those of intermediate luminosity galaxies.  If
there is any discrepancy between the data and models, it lies in the
fact that the observed metallicity distribution function appears to be
dominated by values lower than this.  However, because the
high-metallicity component has a much larger effect on a system's mean
metallicity than its median metallicity, our mean [M/H] value is still
consistent with [M/H] $\gtrsim -1$ (see Figure~\ref{md}, right panel).
Our metallicity result is therefore most consistent with the
simulations of \citet{willman2004} and is somewhat discrepant with
those of \citet{sommer2005}.

\section{Conclusions}

We have used the {\sl ACS\/} to obtain deep F606W and F814W images of
red giant stars in the intracluster space of Virgo about half-way
between M86 and M87.  These images reach to $m_{\rm F814W} \sim 28.5$,
extend $\sim 1.5$~mag down Virgo's red giant branch, and include $\sim
5,300$ intracluster stars.  Careful evaluation and subtraction of the
galaxy background using the UDF demonstrate that the RGB is much wider
than can be explained by our photometric errors.

Fits of these data to theoretical stellar evolution models and
comparisons to ultra-deep surface photometry suggest that Virgo's
intracluster population in this field is dominated by low-metallicity
stars ([M/H]$\lap$-1) with ages $\gap$10~Gyr.  However, the field
appears to contain stars of the full range of metallicities probed
(-2.3$\leq$[M/H]$\leq$0.0).  Our measurements show that there is a
significant number of very metal poor ([M/H]$<-1.5$) stars present in
the field.  Our data also provide an estimate for the amount of solar
metallicity stars; although our F606W observations do not reach the
tip of the very metal-rich RGB, star counts and ultra-deep surface
photometry both constrain this population to be $\lesssim 20\%$.
Stars with [M/H] $>-0.5$ appear to be younger than the rest of the
population, but, interestingly, it is the metal-poor stars that
exhibit more spatial structure.  Overall, the data suggest that
Virgo's intracluster stars are not well-mixed, and have multiple
origins.

The dominant low-metallicity component is a surprise, and the result
could be used to argue that dwarf galaxies are an important source of
intracluster stars.  However, because stars are expected to be removed
from the outer, metal-poor regions of galaxies during tidal
interactions, larger galaxies might also contribute to the
intracluster population.  For example, studies of the scale-lengths of
the disks of spiral galaxies in Coma show that they are smaller than
those of similar galaxies in the field, suggesting that some stars
from the outer parts of the disks are disrupted by harassment
\citep{aguerri2004,gutierrez2004}. In any case, the wide range of
metallicities seen in our data allows for contributions from many
galaxy types.

Our measurement of the metallicity distribution of Virgo's intracluster
stars provides a new constraint for models of the formation and evolution
of galaxy clusters.  Previously this diffuse component was constrained
only by star counts and a very few radial velocity measurements.
Along with the data provided by the four intracluster globular clusters 
present in our field \citep{VICS2}, our results should facilitate the 
development of the next generation of models for the dynamical evolution 
of galaxies in clusters.

Support for this work was provided by NASA through grant number
NAG5-9377 and through grant GO-10131 from the Space Telescope Science
Institute .  We would like to thank Tom Brown for his assistance with
the preparations of the observations.


\clearpage

\begin{deluxetable}{ccc}
\tablecaption{Photometric errors determined from artificial star tests.}
\tableheadfrac{0.01}
\tablehead{
\colhead{{$m_{\rm F814W}$ (Vegamag)}} &
\colhead{{$m_{\rm F814W}$ rms error (mag)}} &
\colhead{{F606W - F814W rms error (mag)}}
}
\startdata
26.0      &           0.03        &       0.03\\
26.5      &           0.04         &       0.04\\
27        &           0.06       &         0.06\\
27.5      &           0.09       &         0.10\\
28        &           0.14       &         0.16\\
28.5      &           0.21      &          0.26\\
29.0      &           0.27      &          0.34\\
\enddata
\label{errors}
\end{deluxetable}

\clearpage

\clearpage

\end{document}